# Genetic Influences on Brain Aging: Analyzing Sex Differences in the UK Biobank using Structural MRI [*]

Karen Ardila[1-3], Aashka Mohite[2-4], Abdoljalil Addeh[1-3], Amanda V. Tyndall[2,3,5], Cindy K. Barha[2,3,6], Quan Long[2,3,5,7], and M. Ethan MacDonald[1-4,8]

[1] Department of Biomedical Engineering, Schulich School of Engineering, University of Calgary, Canada
[2] Hotchkiss Brain Institute, Cumming School of Medicine, University of Calgary, Canada
[3] Alberta Children's Hospital Research Institute, University of Calgary, Canada
[4] Department of Electrical & Software Engineering, Schulich School of Engineering, University of Calgary, Canada
[5] Department of Medical Genetics, Cumming School of Medicine, University of Calgary, Canada
[6] Faculty of Kinesiology, University of Calgary, Canada
[7] Department of Biochemistry & Molecular Biology, Cumming School of Medicine, University of Calgary, Canada
[8] Department of Radiology, Cumming School of Medicine, University of Calgary, Canada

**Synopsis**

**Motivation:** Brain aging varies significantly between sexes, yet genetic contributions to these differences remain under-explored.

**Goal:** Identify sex-specific genetic variants linked to accelerated brain aging using structural MRI data.

**Approach:** This study proposes implementing Brain Age Gap Estimates (BrainAGE) with sex-stratified GWAS to uncover genetic associations in T1-weighted MRI data from the UK Biobank, complemented by Post-GWAS analyses to explore biological pathways and gene expression.

**Results:** Sex-stratified analyses revealed neurotransmitter and mitochondrial response to cellular stress genes linked to brain aging in females and immune-related genes in males. Shared genes suggest common neurostructural roles, advancing understanding of sex-specific genetic determinants in brain aging.

**Impact:** This study highlights the importance of sex-stratified analysis in understanding the genetic associations with brain aging. Findings pave the way for future work on personalized treatments and preventative measures for neurodegeneration based on individual genetic profiles and sex-specific risks.

**Introduction**

Understanding genetic contributions to brain aging and sex differences is crucial for addressing the projected increase in neurodegenerative diseases [1]. Although research has documented different aging trajectories between sexes [2], the specific genetic factors driving these differences in brain aging are less understood. Genetic variants involved in neuroinflammation [3], synaptic integrity [4], and brain morphology [5] are known to contribute to brain aging, but the extent to which these influences vary by sex remains underexplored [6]. The brain age gap estimate (BrainAGE), derived from structural MRI, serves as a valuable biomarker for detecting deviations in brain aging patterns [7]. Leveraging data from the UK Biobank [8], a large-scale genetic and neuroimaging database, provides a robust platform for examining sex-specific genetic influences on brain aging.

This study aims to identify genetic variants associated with brain aging that differ by sex, using functional enrichment analysis to reveal distinct biological processes influencing male and female brain aging. These insights could inform targeted interventions to reduce neurodegenerative risk by addressing sex-specific mechanisms.

**Methods**

The study was conducted using T1-weighted structural MRI and genotyping data from the UK Biobank. The cohort included 40,940 participants aged 45 to 83 years (mean age: 65.21 years, SD: 7.50), with 54.30% female as shown in Figure 1. BrainAGE was calculated by subtracting chronological age from model-predicted brain age based on the features total brain volume (TBV), lateral ventricular volume (LVV), and total hippocampal volume (THV), as shown in Figure 2.

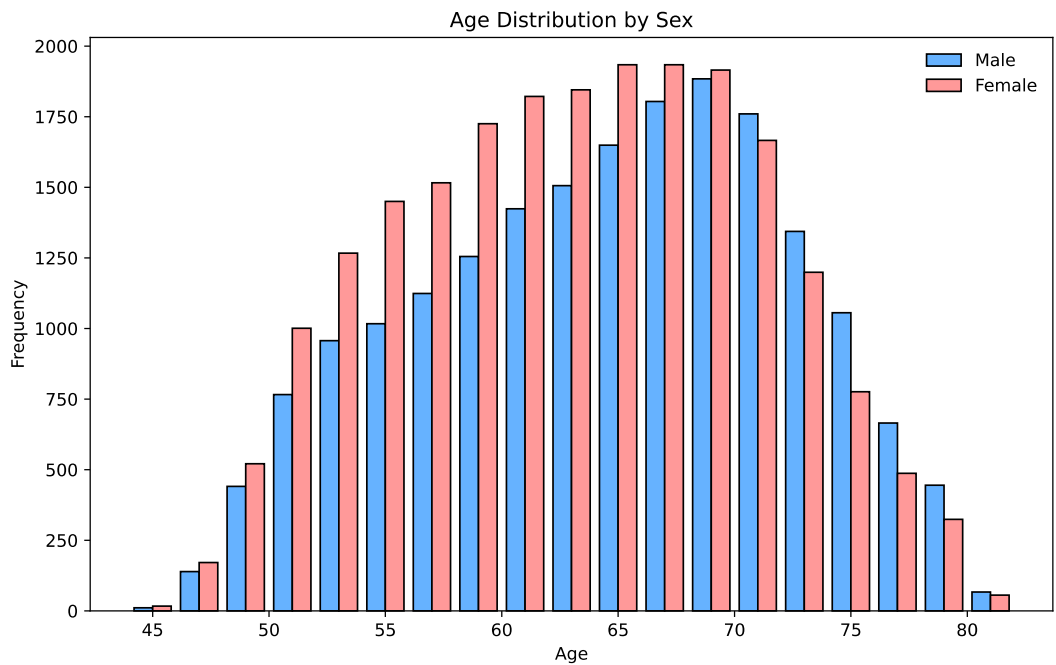

**Figure 1.** Age distribution of the cohort participants by sex. The dataset indicates that both sexes are represented across the age spectrum of 45 to 83 years. There is a higher frequency of females in the younger age groups (45-60 years, mean: 63.71, standard deviation: 7.58) and a slightly higher frequency of males in the older age groups (60-70 years, mean: 65.03, standard deviation: 7.82).

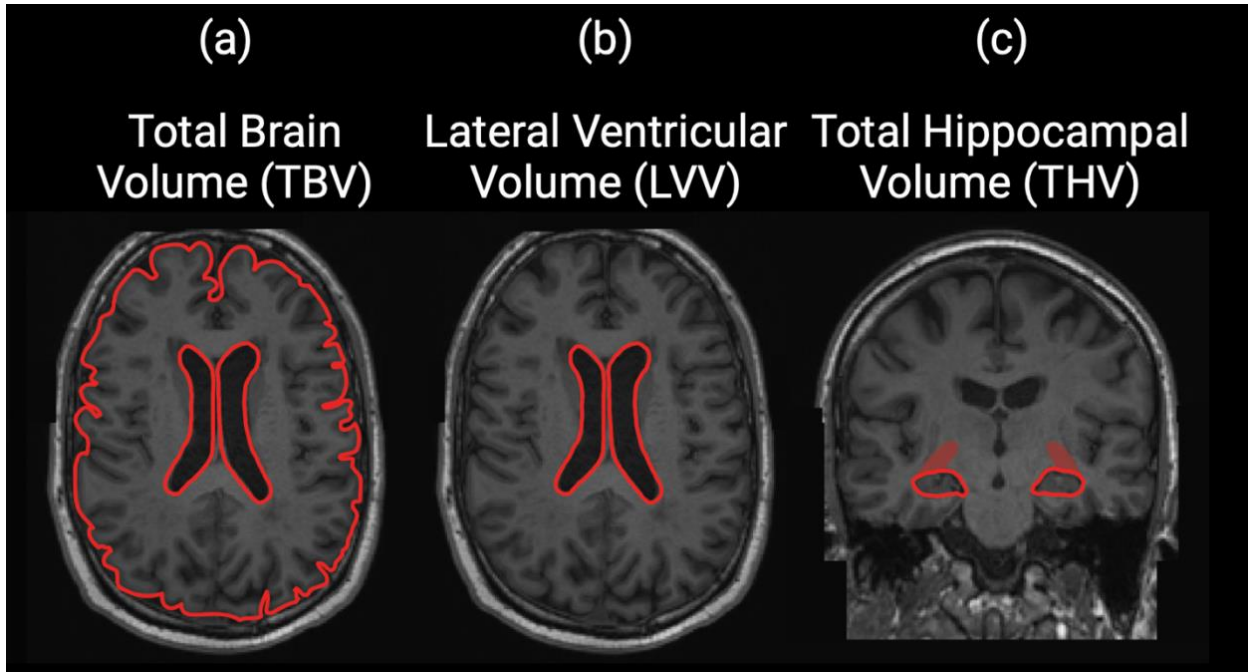

**Figure 2.** T1-weighted MRI-derived regions of interest (ROIs) analyzed: a) Total Brain Volume (TBV), representing the combined volume of grey and white matter, normalized for head size. (b) Lateral Ventricular Volume (LVV), indicating cerebrospinal fluid volume in the lateral ventricles, normalized for head size. (c) Total Hippocampal Volume (THV), encompassing grey and white matter in both hippocampi. Data used in this figure are derived from the UK Biobank resource under application number 55274.

Genome-wide association studies (GWAS) [9-11] were conducted separately for males and females to identify significant single-nucleotide polymorphisms (SNPs) associated with BrainAGE, using a candidate SNP p-value cutoff of 0.05 and a lead SNP p-value threshold of 5e-8. Results were processed in FUMA [12], [13] for visualization and downstream Post-GWAS analysis to explore biological pathways, tissue-specific expression profiles, and gene set enrichment [14].

Post-GWAS analyses were performed using GTEx v8 datasets [15] (54 and 30 tissue types) and the BrainSpan dataset [16] (29 ages and 11 general brain developmental stages). Separate functional enrichment analyses examined biological pathways (KEGG [17], Reactome [18], and such) and gene sets (GO [19], [20], MSigDB [21]). A schematic overview of the methods is shown in Figure 3.

## Results

The BrainAGE models showed robust predictive performance across the brain structures analyzed, as shown in Figure 4. For TBV, the model achieved a mean absolute error (MAE) of 3.18 years and an $R^2$ of 0.74. LVV had a MAE of 3.01 years and an $R^2$ of 0.76, while THV demonstrated the highest accuracy with a MAE of 2.36 years and an $R^2$ of 0.85.

The GWAS analysis highlighted both shared and sex-specific genetic associations with BrainAGE across key brain structures. In females, unique associations included genes such as SLC6A20 [22] and WARS2 [23] for TBV, implicated in neurotransmitter transport and mitochondrial response to cellular stress, roles that may affect neurodegenerative vulnerability by impacting cellular homeostasis. Additionally, TIE1 [24] and SPTBN1 [25] were associated with LVV in females, potentially influencing vascular stability and cytoskeletal structure, thereby indicating sex-specific susceptibilities to cerebrovascular and neurodegenerative changes.

In males, associations with TBV included TGFBR2 [26], involved in cellular signaling and inflammatory response pathways, potentially supporting neuroprotective mechanisms against immune-related neurodegeneration. Shared genetic associations across sexes included WNT16 [27], TNFRSF11B [28], and KCNJ16 [29] in TBV, suggesting foundational roles in neurostructural maintenance.

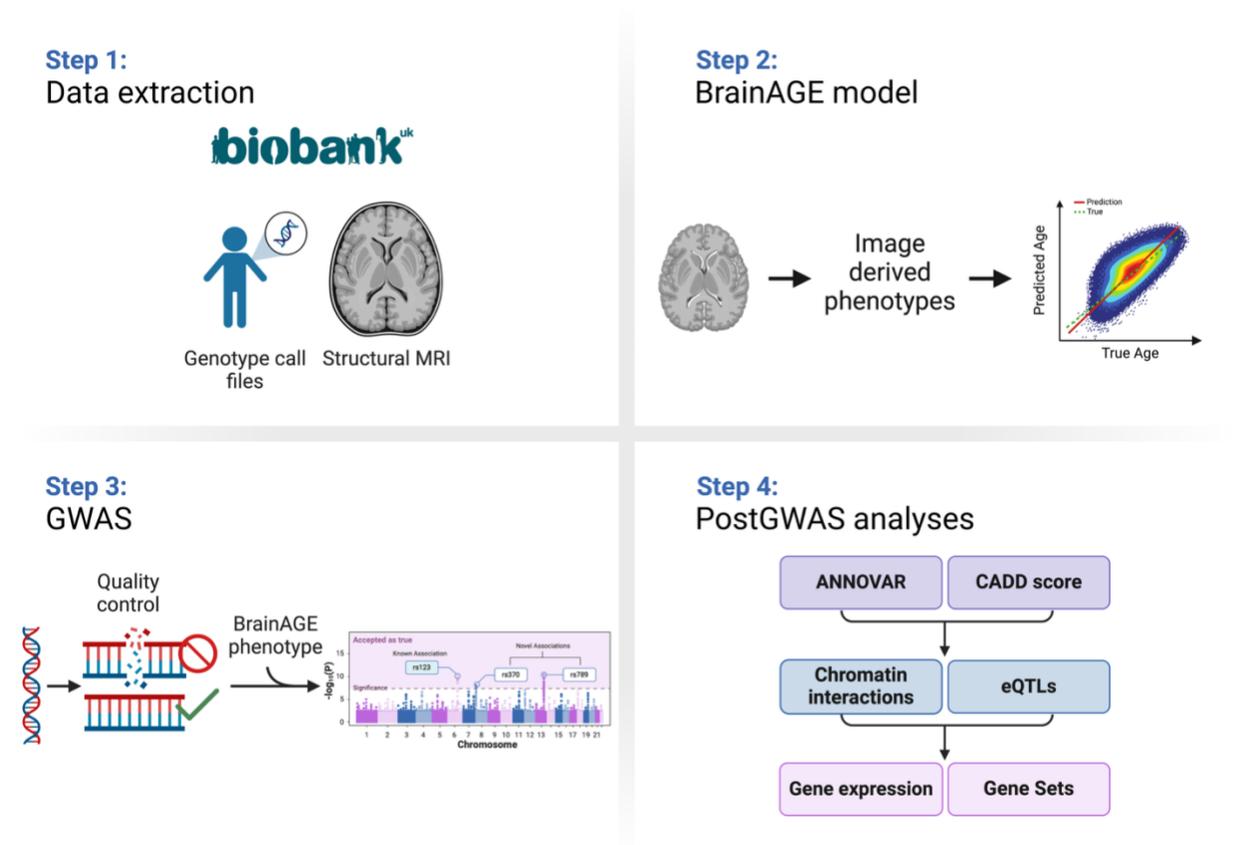

**Figure 3.** Pipeline of the analysis performed. Step 1: Extract and curate neuroimaging and genetic data from the UK Biobank. Step 2: Calculate BrainAGE using MRI-derived volumes from FreeSurfer measurements with a random forest machine learning algorithm and bias correction method. Step 3: Perform quality control with PLINK, conducting GWAS with BrainAGE as the phenotype, visualized in the tool FUMA. Step 4: Conduct PostGWAS analyses in FUMA, including chromatin interactions, tissue-specific expression heatmaps, and gene set enrichment.

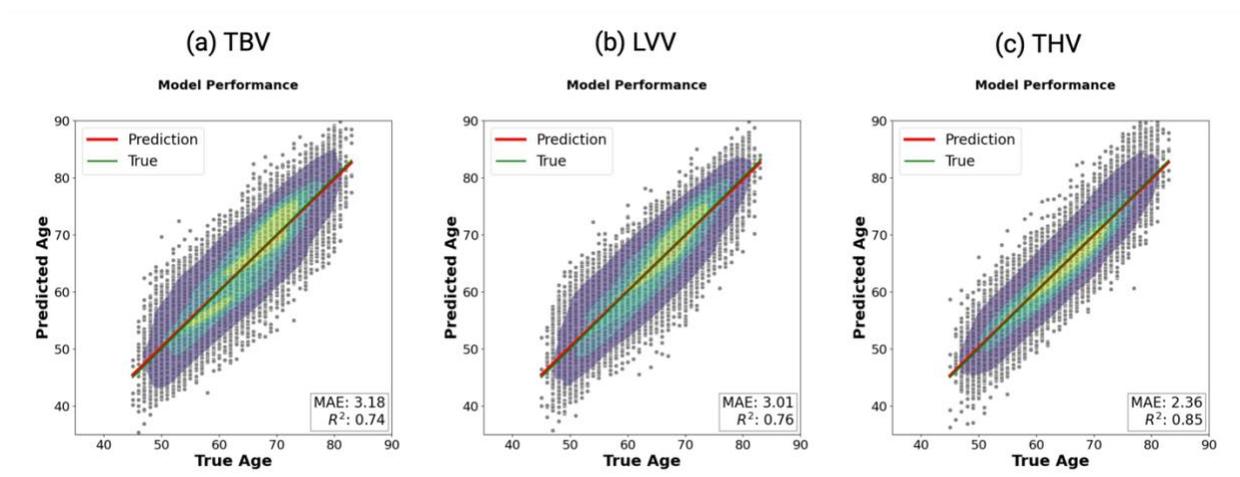

**Figure 4.** Scatter density plot of predicted age versus chronological age for the three regions of interest (ROIs) using BrainAGE model for T1-weighted MRI. (a) Total Brain Volume (TBV), (b) Lateral Ventricular Volume (LVV), and (c) Total Hippocampal Volume (THV). Each plot shows the distribution and correlation of predicted brain age with actual chronological age, highlighting the R-squared score and mean absolute error (MAE) of the BrainAGE model for each ROI.

Genes GMNC [30] and OSTN [31] were identified in multiple analyses (TBV in females, LVV and THV across both sexes), suggesting broad yet differentiated influences on brain volumes. These consistent associations across regions and sexes suggest core regulatory roles that may interact with sex-specific factors in brain aging pathways.

Additional shared genes for LVV, such as PALLD [32], AMZ1 [33], LPAR1 [34], NUAK1 [35], and C16orf95 [36], highlight contributions to ventricular regulation and cerebrospinal fluid balance, essential for structural stability in aging brains. For THV, genes such as FOXO3 [37] and MSRB3 [38], present across sexes, underscore roles in hippocampal maintenance relevant to cognitive resilience, as shown in Figure 5.

Enrichment analyses highlighted distinctions in gene-tissue expression heatmaps relevant to neurodegenerative susceptibility. Certain male-specific genes showed involvement in pathways linked to immune regulation, suggesting roles in neuroinflammatory response, while female-specific genes were associated with cellular stress response pathways, including mitochondrial and oxidative stress mechanisms.

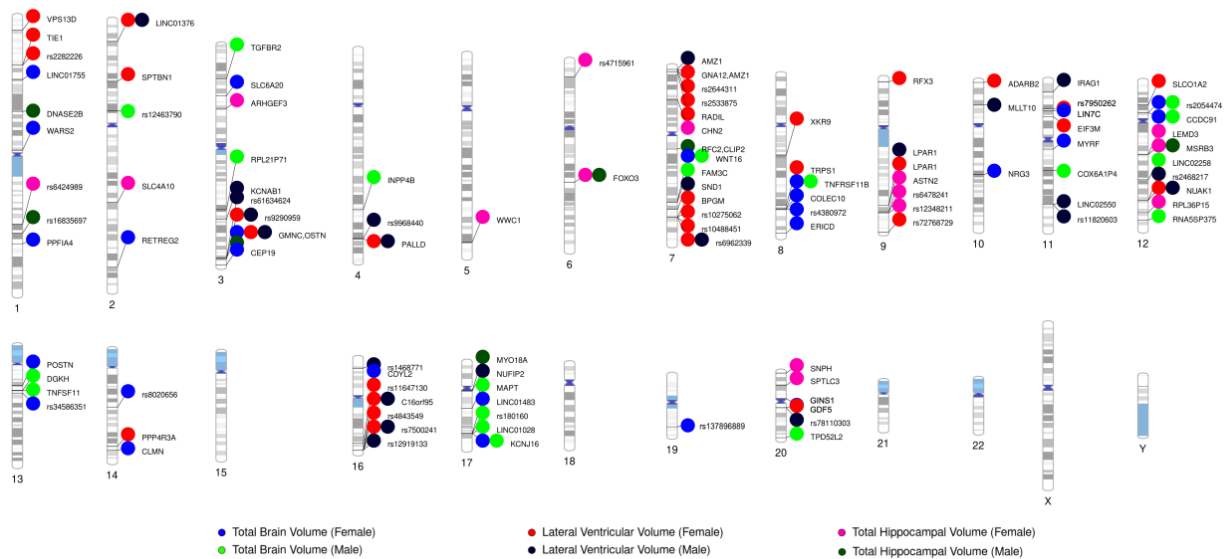

**Figure 5.** Phenogram illustrating chromosomal locations of SNPs associated with BrainAGE in sex-stratified GWAS. Highlighted loci reveal significant SNPs involved in neuroinflammation and brain structure, enhancing understanding of genetic factors underlying sex-specific brain aging dynamics (http://visualization.ritchielab.org/).

## Discussion

This study identified sex-specific genetic influences on brain aging, with implications for tailored neurodegeneration interventions. Distinct genetic patterns observed in males and females support the need for sex-stratified approaches to studying brain aging, possibly due to hormonal influences and genetic expression patterns. Beyond genetic associations, the findings underscore BrainAGE's potential to inform preventative strategies in clinical settings. By integrating genetic and MRI data, this work contributes to precision medicine efforts aimed at mitigating age-related neurological decline across populations.

## Acknowledgements

This research has been conducted using the UK Biobank Resource under Application Number 55274. The authors would like to thank the University of Calgary, in particular the Schulich School of Engineering and Departments of Biomedical Engineering and Electrical & Software Engineering; the Cumming School of Medicine and the Department of Medical Genetics; as well as the Hotchkiss Brain Institute, Research Computing Services (ARC) and the Digital Alliance of Canada for providing resources. KA acknowledges funding from the graduate scholarships Alberta Innovates, Alberta Graduate Excellence Scholarship, Natural Sciences and Engineering Research Council

Brain Create, and Mitacs Globalink Graduate Fellowship. MEM acknowledges support from Start-up funding at UCalgary and a Natural Sciences and Engineering Research Council Discovery Grant (RGPIN-03552) and Early Career Researcher Supplement (DGECR-00124).